\documentstyle[preprint,psfig,prd,aps]{revtex}

\newcommand{\be}{\begin{eqnarray}}
\newcommand{\ee}{\end{eqnarray}}
\newcommand{\no}{\nonumber}

\newcommand{\tr}{\triangle}
\newcommand{\trv}{\vec{\triangle}}
\newcommand{\Lag}{{\cal L}}
\newcommand{\dsla}{D \hspace{-0.2cm}/~}

\newcommand{\x}{{\mbox{$\vec{x}$}}}
\newcommand{\y}{{\mbox{$\vec{y}$}}}
\newcommand{\sig}{{\mbox{$\vec{\sigma}$}}}
\newcommand{\B}{{\mbox{$\vec{B}$}}}
\newcommand{\E}{{\mbox{$\vec{E}$}}}
\newcommand{\gam}{{\mbox{$\vec{\gamma}$}}}
\newcommand{\Sig}{{\mbox{$\vec{\Sigma}$}}}
\newcommand{\Op}{{\cal O}}
\newcommand{\Z}{{\cal Z}}
\newcommand{\e}{{\cal E}}
\newcommand{\f}{{\mbox{$f_B$}}}
\newcommand{\m}{{\mbox{$m_{Q}^{2}$}}}
\def\simgt{\rlap{\lower 3.5 pt\hbox{$\mathchar \sim$}}
           \raise 1pt \hbox {$>$}}
\def\simlt{\rlap{\lower 3.5 pt\hbox{$\mathchar \sim$}}
           \raise 1pt \hbox {$<$}}
\begin{document}
\draft
\preprint{\vbox{\hbox{HUPD-9709}
                \hbox{KEK Preprint 97-60}}}

%\begin{flushright}
%\begin{tabular}{l}
%HUPD-9709 \hspace{1em} \\
%KEK Preprint 97-60
%\end{tabular}
%\end{flushright}

\title{$f_B$ with lattice NRQCD including $O(1/m_Q^2)$ corrections}
\author{K-I. Ishikawa, H. Matsufuru, T. Onogi and N. Yamada \\ 
  {\it Department of Physics, Hiroshima University,} \\
  {\it Higashi-Hiroshima 739, Japan } 
  \vspace{0.2cm} \\
  S. Hashimoto  \\
  {\it High Energy Accelerator Research Organization(KEK)},\\ 
  {\it Tsukuba 305, Japan } }
\maketitle
%%%%%%%%%%%%%%%%%%%%%%%%%%%%%%%%%%%%%%%%%%%%%%%%%%%%%%%%%%%%%%%%%%%
\begin{abstract}
\setlength{\baselineskip}{1ex}
  We calculate the 
  heavy-light meson
  decay constant using lattice NRQCD action for the heavy quark
  and Wilson quark action for the light quark 
  over a wide range in the heavy quark mass. 
  Simulations are carried out 
  on a $16^3 \times 32$ 
  lattice with 
  120 quenched gauge configurations generated with the plaquette action 
  at $\beta=5.8$. 
  For the heavy quark part of the calculation, two sets of
  lattice NRQCD action and current operator are employed.
  The first set includes terms up to $O(1/m_Q)$ both in the action 
  and the current operator, and the second set 
  up to $O(1/m_Q^2)$, where $m_Q$ is the bare mass of the heavy
  quark.  Tree-level values with tadpole improvement are employed 
  for the coefficients in the expansion. 
  We compare the results obtained from the two sets in detail
  and find that the truncation error of higher order
  relativistic corrections for the decay constant are
  adequately small around the mass of the $b$ quark. 
  We also calculate the 1S hyperfine splitting of $B$ meson,
  $M_{B_s} - M_B$ and $f_{B_s}/f_B$ with both sets and find
  that the $1/m_Q^2$ corrections are negligible.
  Remaining systematic errors and the limitation of NRQCD
  theory are discussed.
\end{abstract}
\vspace{2mm}

%%%%%%%%%%%%%%%%%%%%%%% INTRODUCTION %%%%%%%%%%%%%%%%%%%%%%%%%%%%%%%
\setlength{\baselineskip}{0.5ex}

\section{Introduction}

The properties of hadrons including a heavy quark,
particularly $b$ quark, provide us with crucial information 
for constraining the Cabibbo-Kobayashi-Maskawa (CKM)
mixing matrix of the Standard Model, 
which still have large uncertainties in spite of much effort with various 
approaches.
For the combination $|V_{tb}^*V_{td}|$ the current 
value is $0.009 \pm 0.003$\cite{PDG}. 
The large error mainly arises from uncertainties in the decay constant 
$f_B$ and the bag parameter $B_B$ of B meson, which are needed to relate 
the experimentally measured $B^{0}-\bar{B}^0$
transition rates with $|V_{tb}^*V_{td}|$.
It is, therefore, very important for the verification of the 
Standard Model to determine these B meson matrix elements with higher 
accuracy. 

The lattice technique enables us to carry out this task 
from the first principle of Quantum Chromodynamics (QCD). 
In this paper we concentrate on the decay constant 
and study various uncertainties in the calculation, which is 
also instructive for the calculation of the bag parameter. 
Extensive effort has been devoted to a lattice QCD determination of
$f_{B}$ in the past\cite{fb_review}.  
We are now at the second stage where the accuracy 
has become the main issue.
The largest obstacle for obtaining a reliable
prediction is the large value of the $b$ quark mass.
A naive application of the Wilson or $O(a)$-improved fermion 
action for the $b$ quark causes a systematic error of
$O(am_{b})$ in B meson quantities where $m_b$ is the $b$ quark mass and $a$ 
the lattice spacing.
Since $am_{b}$ exceeds unity for lattice
parameters currently accessible in numerical simulations, 
the error is expected to be large, rendering 
an extrapolation to the continuum limit unreliable. 

Non-relativistic QCD (NRQCD) \cite{NRQCD} is designed to
remove the mass scale $m_Q$ of the heavy quark from the theory and
there are no $O(am_{Q})$ systematic errors in this approach.
Since NRQCD is organized as a systematic $1/m_{Q}$ expansion 
of the full relativistic QCD, relativistic errors in NRQCD are 
induced only by the truncation of the $1/m_{Q}$ expansion.
It is, therefore, possible to improve the approximation and
to estimate the remaining uncertainty in a systematic way
based on the $1/m_{Q}$ expansion.

Exploratory studies of the decay constant with lattice
NRQCD were made by Davies {\it et al.} \cite{fb1} and
Hashimoto\cite{fb2}, where only a part of $1/m_{Q}$ terms
was included.
A study with lattice NRQCD action fully including effects
of the $1/m_Q$ terms was carried out in Ref.\cite{NRfB}.  
It was concluded that the magnitude of $O(1/m_Q)$ correction is 
significantly larger than the naive
expectation $\sim O(\Lambda_{QCD}/M_P)$, where $M_P$ is a
pseudo-scalar meson mass, and therefore it was necessary to investigate 
the next order term in NRQCD.  

The goal of the present work is to estimate the magnitude of
higher order $O(1/m_Q^2)$ effects on the $B$ meson decay
constant. 
For this purpose we compare simulation results of the two
sets of the lattice NRQCD action and the operator: 
the first set includes terms up to $O(1/m_Q)$ consistently and the
second set takes into account the entire correction up to $O(1/m_Q^2)$.
Tree-level values with tadpole improvement are employed for 
the coefficients of the correction terms.
We find that the contributions of second order in $1/m_Q$ to
the decay constants is adequately small around the $b$
quark.  
In order to check the generality of the above statement,  
the 1S hyperfine splitting of B meson,  $M_{B_s} - M_B$ and
$f_{B_s}/f_B$ are also investigated and similar results are
obtained. 
Examination of systematic uncertainties other than the relativistic
correction, such as discretization error, quenched approximation and
the one-loop renormalization parameters, are not considered
in detail in this work, leaving them for future studies.
A preliminary report of an investigation of $O(1/m_Q^2)$ 
corrections similar to our work has been reported in Ref.\cite{fbm2}.

This paper is organized as follows.
In section \ref{sec:Lattice_NRQCD} we introduce the action
and the current operators used in our calculation of the
decay constants. 
Simulation details such as parameter values and methods 
are given  in Section \ref{sec:Simulations_and_Results}, followed 
by presentation of results for the decay constant and related 
quantities.
In section \ref{sec:Discussion} implications of the results
and other systematic errors are discussed.
Our conclusions are given in section \ref{sec:Conclusion}. 

%%%%%%%%%%%%%%%%%%%%% lattice NRQCD %%%%%%%%%%%%%%%%%%%%%%%%%
\section{Lattice NRQCD}
\label{sec:Lattice_NRQCD}

NRQCD action at the tree level is obtained from a
relativistic action by Foldy-Wouthuysen-Tani(FWT)
transformation, 
\be
\Lag = \overline{h}(i\dsla - m_Q)h 
\hspace{0.2cm}
\Longrightarrow
\hspace{0.2cm}
\Lag_{\mbox{\small\it NRQCD}} = \Lag_Q + \Lag_{\chi},
\ee
where $h$ is a 4-component spinor of the heavy quark field
and $Q$ and $\chi$ are 2-component fields in the NRQCD
theory. 
The NRQCD action is represented by the following $1/m_Q$ expansion,
\be
\Lag_Q &=& \Lag_Q^{(0)}+\Lag_Q^{(1)}+\Lag_Q^{(2)}+\cdots \no ,\\
\Lag_Q^{(i)}&=& \left( \frac{1}{m_Q} \right)^i Q^{\dag}L^{(i)}Q, 
\ee
where the mass term is discarded since it only amounts to a 
constant shift in the total meson energy and does not affect 
the dynamics of the system.
Lattice NRQCD action is a discretized version of the continuum 
action Wick-rotated to the Euclidean formalism.
The discretization procedure is not unique, 
and we choose a form which leads to the
following evolution equation for the heavy quark
propagator:
\be
G_Q(t,\x) &=& 0  \hspace{3cm} \mbox{(for } t<0)   \\ 
G_Q(t,\x) &=& 
   \left( 1 - \frac{aH_0}{2n} \right)^n 
   \left( 1 - \frac{a\delta H}{2} \right) 
  U_4^{\dag}
   \left( 1 - \frac{a\delta H}{2} \right) 
   \left( 1 - \frac{aH_0}{2n} \right)^n 
   G_Q(t-1,\x)                         \no \\ 
&& + \delta_{x,0}  \hspace{2.3cm} \mbox{(for } t\ge0).
\ee
Here $x=(t,\x)$, $n$ is the stabilizing parameter\cite{NRQCD}.
Our discretization procedure is almost same as 
\be
G_Q(t,\x) =
\left( 1 - \frac{a\delta H}{2} \right)
\left( 1 - \frac{aH_0}{2n} \right)^n 
U_4^{\dag}
\left( 1 - \frac{aH_0}{2n} \right)^n 
\left( 1 - \frac{a\delta H}{2} \right) 
G_Q(t-1,\x), \no
\ee
which was used in \cite{fbm2}.
These two discretization procedures are the best choices from the view
of the control on the discretization error in the temporal derivative.

$H_0$ and $\delta H$ are defined as follows:
\be
&&H_0            = - \frac{\triangle^{(2)}}{2m_Q},       \\
&&\delta H       = \sum_i c_i \delta H^{(i)},            \\
&&\delta H^{(1)} = - \frac{g}{2m_Q}\sig \cdot \B,        \\
&&\delta H^{(2)} =   \frac{ig}{8m_Q^2}
                     ( \trv \cdot \E - \E \cdot \trv ),  \\
&&\delta H^{(3)} = - \frac{g}{8m_Q^2}\sig \cdot 
                     (\trv \times \E - \E \times \trv ), \\
&&\delta H^{(4)} = - \frac{(\triangle^{(2)})^2}{8m_Q^3} ,\\
&&\delta H^{(5)} =   \frac{a^2 \tr^{(4)}}{24m_Q},        \\
&&\delta H^{(6)} = - \frac{a (\tr^{(2)})^2}{16nm_Q^2}.
\ee
The symbols $\trv$ and $\tr^{(2)}$ denote the symmetric lattice
differentiation in spatial directions and Laplacian,
respectively, and $\tr^{(4)} \equiv \sum_i (\tr^{(2)}_i)^2$.
The field strengths $\B$ and $\E$ are generated from the standard 
clover-leaf operator.

The coefficients $c_i$ in (6) should be determined by perturbatively
matching the action to that in relativistic QCD. 
In the present work we adopt the tree-level value $c_i=1$ for all $i$ and 
apply the tadpole improvement\cite{MFimp} 
to all link variables in the evolution equation by
rescaling the link variables as $U_{\mu} \rightarrow
U_{\mu}/u_0$. 
The value of $u_0$ is given in Section \ref{subsec:Parameters}.

The original 4-component heavy quark spinor $h$
is decomposed into two 2-component spinors $Q$ and $\chi$
after FWT transformation,
\be
h(x) = R \left( \begin{array}{c} Q(x) \\ \chi^{\dag}(x) 
                    \end{array} \right),
\ee
where $R$ is an inverse FWT transformation matrix which has
$4\times4$ spin and $3\times3$ color indices.
After discretization, $R$ at the tree level is written as
follows: 
\be
&&R       = \sum_{i}R^{(i)},          \\
\label{eq:R(1)}
&&R^{(1)} =   1 ,                           \\
\label{eq:R(2)}
&&R^{(2)} = - \frac{\gam\cdot\trv}{2m_Q}  , \\             
\label{eq:R(3)}
&&R^{(3)} =  \frac{\tr^{(2)}}{8m_Q^2}     , \\
\label{eq:R(4)}
&&R^{(4)} =   \frac{g\Sig\cdot\B}{8m_Q^2} , \\ 
\label{eq:R(5)}
&&R^{(5)} = - \frac{ig\gamma_4\gam\cdot\E}{4m_Q^2},   
\ee
where
\be
\Sigma^{j} = \left( \begin{array}{cc} \sigma^{j} &     0      \\
            0      & \sigma^{j} \\ \end{array} \right).
       \ee
The tadpole improvement\cite{MFimp} is also applied for these 
operators in our simulations. 

As mentioned in the Introduction, we define two sets of
action and FWT transformation  \{$\delta H$,$R$\} as follows: 
\be 
\mbox{set\ I} \equiv \{ \delta H_1,R_1 \}
\hspace{0.5cm}{\rm and}\hspace{0.5cm} 
\mbox{set\ I$\!$I} \equiv \{ \delta H_2,R_2 \},
\ee
where
\be
\delta H_1 = \delta H^{(1)} 
\hspace{0.5cm} &{\rm and}& \hspace{0.5cm} 
R_1 = \sum_{i=1}^{2}R^{(i)},                 \\
\delta H_2 = \sum_{i=1}^{6}\delta H^{(i)}
\hspace{0.5cm} &{\rm and}& \hspace{0.5cm} 
R_2 = \sum_{i=1}^{5}R^{(i)}.
\ee
The operators 
$\delta H_1$ and $R_1$ keep only $O(1/m_Q)$ terms while
$\delta H_2$ and $R_2$ include the entire $O(1/m_Q^2)$ terms and
the leading relativistic correction to the dispersion
relation, which is an $O(1/m_Q^3)$ term.
The terms improving the discretization errors appearing in
$H_0$ and time evolution are also included.
Using these two sets, we can realize the level of accuracy of $O(1/m_Q)$ 
and $O(1/m_Q^2)$ for the set I and II.

%%%%%%%%%%%%%%%%%%%%% SIMULATIONS and Results%%%%%%%%%%%%%%%%%%%%%%%%%
\section{Simulations and Results}
\label{sec:Simulations_and_Results}

\subsection{Parameters}
\label{subsec:Parameters}

Our numerical simulation is carried out with 120 quenched
configurations on a $16^3 \times 32$ lattice at $\beta = 5.8$.  
Each configuration is separated by $2,000$ pseudo-heat bath
sweeps after $20,000$ sweeps for thermalization and fixed to
Coulomb gauge. 
For the tadpole factor we employ 
$u_0 = \langle P_{plaq} \rangle^{1/4}$ with 
$P_{plaq}$ the average plaquette, which takes the value
$u_0 = 0.867994(13)$ measured during our configuration generation. 

For the light quark we use the Wilson quark action with
$\kappa$=0.1570, 0.1585 and 0.1600, imposing the periodic and
Dirichlet boundary condition for spatial and temporal
directions, respectively. 
The chiral limit is reached  at $\kappa_c = 0.16346(7)$ and 
the inverse lattice spacing determined from the rho meson mass
equals $a^{-1} = 1.714(63)$ GeV.
The hopping parameter $\kappa_s$ corresponding to the strange quark is
determined in two ways from $m_{\phi}/m_{\rho}$ and $m_{K}/m_{\rho}$, 
which yields $\kappa_s = 0.15922(39)$ and $0.16016(23)$, respectively. 
In our analysis we take $\kappa_s=0.1600$ for simplicity,
except in the final results where the error arising from 
the uncertainty in $\kappa_{s}$ is taken into account.
We use the factor $\sqrt{1-\frac{3\kappa}{4\kappa_c}}$ as
the field normalization for light quark\cite{KLM}.

For the heavy quark part of calculations, two sets of
lattice NRQCD action and current operator are employed as
described in Section \ref{sec:Lattice_NRQCD}.
For the heavy quark mass and the stabilizing parameter, we
use $(am_Q,n)$=(5.0,2), (2.6,2), (2.1,3), (1.5,3), (1.2,3)
and (0.9,4), which cover a mass range between $2m_b$ and
$m_c$. 

All of our errors are estimated by a single elimination jack-knife
procedure. 

%%%%%%%%%%%%%%%%%%%%%%%%%%%%%%%%%%%%%%%%%%%%%%%%%%%%%%%%%%%%%%%%%%%
\subsection{Method}
\label{subsec:Method}

In the continuum the pseudo-scalar and vector meson decay
constants are defined by
\be
\langle 0 |A_0| P   \rangle &=& f_PM_P ,              \\
\langle 0 |V_i| V_i \rangle &=& \epsilon_i f_VM_V    ,
\ee
where $A_0=\overline{q}\gamma_5\gamma_4h$ and 
$V_i=\overline{q}\gamma_ih$.

The lattice counterpart is calculated in the following way.
Let us define an interpolating field operator for 
heavy-light meson from a light anti-quark and
a heavy quark field by 
\be
  \Op_{X}^{\mbox{\small\it src}}(x) 
= \sum_{\y} \overline{q}(x) \Gamma_X 
  \left(\begin{array}{c} Q(t,\y) \\ 0
        \end{array} \right) \phi^{\mbox{\small\it src}}(|\x-\y|),
\ee
where $\Gamma_X$ is the gamma matrix specifying the quantum
number of the meson. 
The subscript $X$ labels the pseudo-scalar meson ($P$) or
the vector meson ($V$), $\phi$ is a source function and
the superscript `$\mbox{\it src}$' denotes the choice of 
smearing, {\it i.e.,} `$L$'(local) or `$S$'(smeared), according to
\be
\phi^{L}(x) = \delta(x)
\hspace{0.5cm} {\rm or} \hspace{0.5cm}
\phi^{S}(x) = \exp( -a|\x|^b)
\ee
where $a$ and $b$ are fixed by a fit to the Coulomb gauge  wave function
measured in the simulation.
We next define the local axial-vector and vector
currents $J_X$:
\be
    J_{X} 
&=& \overline{q}(x) \Gamma_X h(x)  \no\\ 
&=& \sum_{i} J_X^{(i)}
 =  \sum_{i}\overline{q}(x) \Gamma_X R^{(i)}
    \left( \begin{array}{c} Q(x) \\ \chi^{\dag}(x) \end{array} \right), 
\ee
with
\be
\Gamma_P = \gamma_5\gamma_4 \ \ , \ \ \Gamma_V = \gamma_i.
\ee
The inverse FWT transformation $R^{(i)}$s are explicitly
written in eq.(\ref{eq:R(1)})--(\ref{eq:R(5)}).

To extract the decay constant of heavy-light mesons, we
calculate the following two point functions:
\be
    C_{\Op_X}^{L}(t) 
&=& \sum_{\x}\langle \Op_{X}^{L}(t,\x) 
         \Op_{X}^{L \dag}(0)\rangle , \\ 
    C_{\Op_X}^{S}(t) 
&=& \sum_{\x}\langle \Op_{X}^{L}(t,\x) 
         \Op_{X}^{S \dag}(0)\rangle , \\ 
    C_{J_X}^{S (i)}(t) 
&=& \sum_{\x}\langle J_{X}^{(i)}(t,\x) 
         \Op_{X}^{S \dag}(0)\rangle. 
\ee
We show the effective mass plot of pseudo-scalar meson at
$am_Q=2.6$ and $\kappa = 0.1585$ with local
$C_{\Op_P}^{L}(t)$ and smeared source $C_{\Op_P}^{S}(t)$ for
the set II in Fig.\ref{efplt1}. 
We observe that the effective mass for the smeared source
is very stable from early time slices.
From inspection of effective mass plots such as Fig.\ref{efplt1}, 
we conclude that the ground state of 
pseudo-scalar meson is sufficiently isolated
in the range $[t_{min},t_{max}]=[17,22]$ for both correlators,
and we adopt this range as our fitting interval.

Similar plots for $C_{J_P}^{S (i)}(t)$ (i=2,3,4,5) with the same 
parameters are shown in Fig.\ref{efplt2}.  We find that
the different operators give a consistent value for the ground state
energy.
Effective mass plots for other values of $am_Q$ and $\kappa$ 
exhibit similar features.

In order to extract the binding energy and amplitude, we fit
the correlation functions to the following forms:
\be
    C_{\Op_X}^{S}(t) 
&=& Z_{\Op_X}^{S}exp \left(-E_{\Op_X}^{S}t \right), \\
    C_{\Op_X}^{L}(t) 
&=& Z_{\Op_X}^{L}exp \left(-E_{\Op_X}^{L}t \right), \\
    C_{J_X}^{S (i)}(t) 
&=& Z_{J_X}^{S (i)}exp \left(-E_{J_X}^{S (i)}t \right).
\ee
As is expected from Fig.\ref{efplt1} and Fig.\ref{efplt2}, 
all correlators with the same parameters $am_Q$, $\kappa$
and $X$ give a consistent value of $E$ irrespective of the 
choice of `$L$' or `$S$' and `$\Op$' or `$J$'.  
In the final analysis, we fit the correlators with smeared 
sources to obtain $E_{\Op_X}^{S}$ and fit other correlators 
with $E_{\Op_X}^{S}$ as the input energy. 
We list results for the binding energy $E_{\Op_X}^{S}$ in 
Table \ref{bind2}. 

Chiral extrapolation of $E_{\Op_X}^{S}$ is shown in
Fig.\ref{mas-extrp} and the extrapolated values are
given in the last column of Table \ref{bind2}. 
From Fig.\ref{mas-extrp} we find that the slope become a
little milder for larger $am_Q$.
We also show the $1/(am_Q)$ dependence of binding energy 
for pseudo-scalar meson at $\kappa = \kappa_c$  in
Fig.\ref{Bind}. 

The lattice value of
the meson decay constant is obtained in terms of the amplitudes 
defined above as follows:
\be
f_{X} \sqrt{M_{X}} &=&
a^{-3/2} \sum_{i,j} \Z_{X ij} \frac{\sqrt{2Z_{\Op_X}^L}Z_{J_X}^{S(j)}}
         {Z_{\Op_X}^S} \no\\
&\equiv& a^{-3/2} \sum_{i,j} \Z_{X ij} \delta f_X^{(j)} \no\\
&\equiv& a^{-3/2} (f_X\sqrt{M_X})^{latt},
\ee
where $\delta f^{(j)}_X=\sqrt{2Z_{\Op_X}^L}Z_{J_X}^{S(j)}/Z_{\Op_X}^S$ 
and $(f_X\sqrt{M_X})^{latt}$ are defined for convenience of 
discussions below.
The renormalization constant $\Z_{X ij}$ representing 
mixing effects between different operators are generally needed 
to relate the current operator on the lattice to that in the continuum.
A perturbative calculation for our definition of the NRQCD 
action has not yet been finished, and we treat $\Z_{X ij}$ as a unit
matrix in the present article.
We should remark that the NRQCD Collaboration has recently 
completed such a calculation\cite{tsukuba97}.  
Their choice of action, however, 
is slightly different from ours,  and the results are not applicable 
for our analysis.
Thus our results are stated without the one-loop $Z$-factor, 
but incorporating the mean field improvement.

%%%%%%%%%%%%%%%%%%%%%%%%%%%%%%%%%%%%%%%%%%%%%%%%%%%%%%%%%%%%%%%%%%%%
\subsection{Analysis of results}
\label{subsec:Analysis_at_mean_field_tree_level}

The numerical results of $( f_P \sqrt{M_P} )^{latt}$ for 
all $\kappa$ and $\kappa_c$ are tabulated in Table
\ref{frootm} and its chiral extrapolation is shown in
Fig.\ref{chi-frootm}. 
We find from this figure that the linear extrapolation is
very smooth. 
In contrast to the binding energy, the slope tends to
increase as the heavy quark mass becomes larger. 

Fig. \ref{fmz} shows the $1/(aM_P)$ dependence of 
$( f_P \sqrt{M_P} )^{latt}$ at $\kappa = \kappa_c$ for the set I (open 
symbols) and II (filled symbols), where
$aM_P$ is the pseudo-scalar meson mass in lattice units calculated 
as described below.
Shaded bands represent the mass region corresponding to
the B and D meson.  They are estimated from the value of $a^{-1}$ 
determined from the $\rho$ meson mass
and a string tension($\sim 1.3$GeV) at $\beta=5.8$. 
Solid curves are results of a fit with a quadratic function in
$1/(aM_P)$ given by
\be
  ( f_P \sqrt{M_P} )^{latt} 
= (f_P \sqrt{M_P})^{\infty}
  \left( 1 + \frac{a_1}{aM_P} + \frac{a_2}{(aM_P)^2} \right).
\label{expand}
\ee
The values of the fitted parameters are tabulated in Table \ref{fit}. 
We observe that $(f_P \sqrt{M_P})^{\infty}$ and $a_1$
are consistent between the set I and II within the statistical error,
while $a_2$ is  different as expected.

The meson mass is given by
\be
aM_P = \Z_mam_Q + E_{\Op_P}^{S} - \e_0,
\ee
where $\Z_m$ and $\e_0$ are the mass renormalization factor
and the energy shift, respectively.
Since perturbative results for these quantities are not fully
available for our NRQCD action, we set
$aM_P = am_Q + E_{\Op_P}^{S}$ in this work.
For the action $I$, for which our one-loop
results are available, the one-loop correction is very
small ($\sim$ 5\%) and we expect that effects for the 
final prediction for $f_{B}$ is not significant.

It may appear at first sight that there are no large difference
over almost all of the region of $aM_P$  between the results from set I
(open circles) and those from set II (solid circles) in Fig.\ref{fmz}. 
In order to investigate the differences further, we
decompose the results into contribution of each operator $\delta
f_P^{(i)}$. 
Fig. \ref{fmz_lead} shows the leading contribution 
$\delta f_P^{(1)}$ for each set.
The values from the set II are larger than those from the set I,
showing effects of the $1/m_{Q}^{2}$ correction in the
action. 
In Fig.\ref{fmz_other} the other current contributions  
$\delta f_{P}^{(i)}$ (i=2 for set I and i=2,3,4,5 for
set II) are shown.
As expected, the magnitude of the $O(1/m_Q)$
operator(circles) is larger than the other $O(1/m_Q^2)$
operators(other symbols).
The numerical data for $\delta f_P^{(i)}$ are tabulated in
Table \ref{dfmzP}.

In summary, $O(1/m_Q^2)$ correction in the action tends to raise
$( f_P \sqrt{M_P} )^{latt}$ while the one in the current lower.
After all we find 
a remarkable fact that the small difference in 
Fig.\ref{fmz} results from a cancellation between the 
correction from the action and that from operators, and 
between different operators.

In order to quantify the magnitude of $O(1/m_Q^2)$ corrections, we
define the following quantity,
\be
        \tr f
&\equiv& \frac{   ( f_P\sqrt{M_P} )^{latt}_I 
                - ( f_P\sqrt{M_P} )^{latt}_{I\!I}  }
              {   ( f_P\sqrt{M_P} )^{latt}_I  }, 
\ee
where I and II corresponds to the set I and II.
We show $\tr f$ in Fig.\ref{total_corr}. 
We see that the $O(1/m_Q^2)$ correction is about 3\%
around the B meson region, while increasing to about 15\% 
around the D meson. 
Unless the $O(1/m_Q^2)$ correction in the renormalization factor 
$Z_{A}$ is unexpectedly large, the smallness of
$O(1/m_Q^2)$ corrections in the region of B meson will  be
retained in the physical prediction of the decay constant.  

We have seen that $(f_P\sqrt{M_P})^{latt}$ does not
change much with inclusion of $O(1/m_Q^2)$ terms.  Since  
this is due to a cancellation among the $O(1/m_Q^2)$ contributions
whose origin is not apparent, 
the smallness does not necessarily mean that higher order corrections 
of the $1/m_Q$ expansion are negligible.
To examine this point, 
we define the following quantity:
\be
       |\tr f|
&\equiv& \left(
            |\delta f^{(1)}_{P I}  -  \delta f^{(1)}_{P I\!I}|
          + |\delta f^{(2)}_{P I}  -  \delta f^{(2)}_{P I\!I}|
          + |\delta f^{(3)}_{P I\!I}| \right. \no \\
&&\left.  + |\delta f^{(4)}_{P I\!I}|
          + |\delta f^{(5)}_{P I\!I}|  \right)/
             ( f_{P}\sqrt{M_P} )^{latt}_I.
\ee
This quantity provides a conservative estimate of $O(1/m_Q^2)$
correction since all $O(1/m_Q^2)$ corrections are added.
The $1/(aM_P)$ dependence of $|\tr f|$ is shown in
Fig.\ref{imag_total_corr} with open symbols, together with  the result for 
$\tr f$ (solid symbols).
If we estimate the unknown $O(1/m_Q^3)$ correction to have a magnitude 
$\simlt |\tr f|$, which is presumably an overestimation, we 
deduce that 
 $( f_P\sqrt{M_P})^{latt}_{II}$ 
would be corrected by only about 6\% around B meson. 
On the other hand, there is no reason that $O(1/m_Q^3)$
correction would be small in the D meson region. 

For completeness, we also show the result of the vector meson decay
constant $(f_V\sqrt{M_V})^{latt}$, and the spin average and the ratio of
pseudo-scalar and vector decay constants. 

The numerical results for the vector meson decay constant are
tabulated in Table \ref{dfmzV}.
The results for  $(f_V\sqrt{M_V})^{latt}$ show only small 
difference between the set I and II as in the
pseudo-scalar case.  
Making a decomposition into current components as before, we
find that there are cancellations among $\delta f_V^{(i)}$ as in the
pseudo-scalar, though to a lesser extent.
The numerical data for the spin averaged decay constant
$       ( f\sqrt{M} )_{AV}^{latt} 
\equiv \frac{   ( f_P\sqrt{M_P} )^{latt}
             + 3( f_V\sqrt{M_V} )^{latt} }{4} $ 
can be found in Tab.\ref{n-dfmzAV}. 
The behavior of $( f\sqrt{M} )_{AV}^{latt}$ as a function of 
$aM_{AV}$ is shown in Fig.\ref{fmzAV} where
$aM_{AV}\equiv(aM_P+3aM_V)/4$.

Fig. \ref{PVfig} shows $(f_P/f_V)^{latt}$(circles) and
$(f_P/f_V)^{(1)}$(diamonds) with set I(open symbols) and set
II(solid symbols).
For the numerical data, see Tab.\ref{PVtab}.
%%%%%%%%%%%%%%%%%%%%%%%%%%%%%%%%%%%%%%%%%%%%%%%%%%%%%%%%%%%%%%%%%
\subsection{Other quantities}
\label{subsec:Other_quantities}

In order to find how large the $O(1/m_Q^2)$ corrections are in 
other quantities, we compare 1S hyperfine
splitting, $M_{P_S} - M_P$ and $f_{P_S}/f_P$ obtained with the 
set I and II.
Renormalization effects are expected to be negligible for these 
quantities since the light quark mass dependence of the renormalization 
constant is very small around the chiral limit. 

Fig. \ref{1S-hyper} shows the $aM_P$ dependence of 1S
hyperfine splitting.
The splitting linearly increases with $1/(aM_P)$ for
both sets and $O(1/m_Q^2)$ terms do not affect this
quantity.
The results for $aM_{P_s} - aM_P$ are shown in
Fig.\ref{MBs-MB}. 
We expect from experiments that this quantity depends
only weakly on the heavy quark mass ($M_{B_s^0}-M_{B_d^0}$ =
90.1MeV and $M_{D_s^{\pm}}-M_{D_d^{\pm}}$ = 99.2MeV).  Our results 
are consistent with this expectation including the trend that the 
mass difference increases for smaller heavy quark mass, albeit errors 
are large.
Finally we show $f_{P_s}/f_P$ in Fig.\ref{fBs/fB} calculated 
from the jackknife samples of the following ratio: 
\be
  f_{P_s}/f_P 
= \frac{( f_{P_s}\sqrt{M_{P_s}} )^{latt}}
                   {( f_P\sqrt{M_P} )^{latt}}
  \times \sqrt{\frac{aM_{P}}{aM_{P_s}}}.
\ee
The ratio $f_{B_s}/f_B$ has phenomenological importance since it 
is necessary to extract the Standard Model parameter 
$|V_{ts}|$ which is up to now only poorly determined. 
One can see in Fig.\ref{1S-hyper}-\ref{fBs/fB} that  there is
no significant difference between the results from the two sets
of simulations over almost all mass region up to the D meson. 
Numerical results are tabulated in Table \ref{num-results}. 

With $a^{-1}$ from $\rho$ meson mass, we find $M_P(am_Q=2.6)\sim5.3$GeV
which is close to the experimental value of $M_B$.  
We therefore consider $am_b=2.6$ to be the physical point for the 
$b$ quark and convert the numerical results above  from the set II 
into physical units.
We obtain
\be
M_{B^*}-M_B & = & 26 \pm 9 (statistical)                \,\mbox{MeV},\no \\
M_{B_s}-M_{B_d} & = & 99 \pm 8 (statistical)\pm 13 (strange)\,\mbox{MeV},\no \\
\frac{f_{B_s}}{f_{B_d}} & = & 1.23 \pm 0.03 (statistical)\pm 0.03 (strange).\no\ee
where `$strange$' means the error arising from the 
ambiguity in $\kappa_s$ using $m_{\phi}/m_{\rho}$ or
$m_{\eta}/m_{\rho}$. 
The hyper fine splitting is much smaller than the experimental
value of 46 MeV.
It is known that this quantity is very sensitive to $O(a)$
error and quenching effects.
Other quantities are in reasonable agreement with
experiment and results of previous lattice studies.

%%%%%%%%%%%%%%%%%%%%%%% DISCUSSION %%%%%%%%%%%%%%%%%%%%%%%%%%%%%%%
\section{$\f$ to $O(1/ \m )$ and remaining systematic uncertainties}
\label{sec:Discussion}

Our investigation shows that relativistic corrections of order  
$1/m_{Q}^{2}$ is small in the region of B meson, and that 
higher order corrections are likely to be bound within a 5\% level.
One of the remaining source of systematic uncertainties is 
a discretization error of form  $O((a\Lambda_{QCD})^n)$. 
The leading error of this form existing in our calculation 
is $O(a\Lambda_{QCD})$  which appears from the Wilson
fermion action since the gauge and NRQCD action have no
$O(a)$ term.
The characteristic size of $O(a\Lambda_{QCD})$ at
$\beta$=5.8 is 20--30\%.
This error can be reduced to the level of 5\% by the use of 
$O(a)$-improved Wilson actions.
Alternatively, one may carry out simulations at a larger value of 
$\beta$ in order to reduce the $O(a\Lambda_{QCD})$ error within 
the Wilson action for light quark.
However,  care must be taken in this alternative because
of the problem of divergence of one-loop coefficient
for $m_{Q}\leq$ 0.6--0.8\cite{renorm_NR}.
Such a situation can arise when the heavy quark mass 
parameter in lattice units becomes small, which will be encountered 
in simulations at large $\beta$.
These limitations in the values of $\beta$ and
$1/m_Q$ should be kept in mind in simulations with 
lattice NRQCD.

Another source of systematic errors is the deviation 
of the expansion parameters of the NRQCD action and 
renormalization constants of currents from their tree-level 
values.  Perturbative corrections in these quantities are 
not negligible in general, amounting to 
$\sim$ 20\% at one-loop order.
The renormalization factor of the axial-vector
current in the static limit is known to be particularly 
large, and $1/m_{Q}$ corrections could also be important.
We expect, however, that after including the one-loop correction 
the systematic error of this origin will be reduced to 
$O(\alpha_{s}^{2})\sim$ 5\% in magnitude.

Taking into account the uncertainties discussed above, we quote 
the following estimate from the present work for the B meson decay 
constant in the quenched approximation:
\be
f_B & = & 184 \pm 7(statistical)\pm 5(relativistic)\mbox{ MeV}, \no
\ee
with discretization and perturbative errors of 20 \% each.
We use the static result $Z_A= 1- 0.057 g_V^2(q^{\ast})$ 
\cite{EichtenHill} for the renormalization constant, and an average is 
taken of the result with $q^{\ast}=\frac{1}{a}$ and 
$q^{\ast}=\frac{\pi}{a}$.

%%%%%%%%%%%%%%%%%%%%% CONCLUSION %%%%%%%%%%%%%%%%%%%%%%%%%
\section{Conclusion}
\label{sec:Conclusion}

In this article we have presented a study of the $O(1/m_Q^2)$ correction 
to the decay constant of heavy-light meson with lattice NRQCD and Wilson 
quark action in the quenched approximation.
While the $O(1/m_Q)$ correction to the decay constant in the 
static limit is significant, we find in our
systematic study of $1/m_Q$ expansion that the $O(1/m_Q^2)$
correction is sufficiently small for B meson, so that there will be no need for
incorporating $O(1/m_Q^3)$ corrections unless an accuracy of better than
5\% is sought for.
Our examination of other physical quantities in the same
respect also provides encouraging support to this 
statement. 
We have thus shown using our highly improved lattice NRQCD that the
relativistic error, which has been one of the largest
uncertainty in lattice calculations of the B meson decay constant, 
is well under control. 

Our results still have several sources of large systematic errors. 
In order to obtain $f_B$ with a higher precision, we need to reduce
(i) the $O(a)$ error by using an $O(a)$ improved Wilson fermion action 
for the light quark, 
(ii) the $O(\alpha_s)$ and $O(a \alpha_s)$ errors with 
  the fully one-loop corrected perturbative renormalization
  coefficients for both the action and the operator, and 
(iii) the scale setting and quenching error by
  doing simulations with full QCD configurations. 
The problems (i) and (ii) are currently under study, and  
we are planning to pursue (iii) soon.
When these improvements are all in place, we expect to achieve a lattice 
NRQCD determination of $f_B$ with the accuracy of less than 10\%.

%%%%%%%%%%%%%%%%%%%%% ACKNOWLEDGMENT %%%%%%%%%%%%%%%%%%%%%%%%%
\section{Acknowledgment}

Numerical calculations have been done on Paragon XP/S at
INSAM (Institute for Numerical Simulations and Applied
Mathematics) in Hiroshima University.
We are grateful to S. Hioki for allowing us to use his
program to generate gauge configurations.
We would like to thank J. Shigemitsu, C.T.H. Davies,
J. Sloan, Akira Ukawa and the members of JLQCD collaboration for 
useful discussions.
H.M. would like to thank the Japan Society for the Promotion
of Science for Young Scientists for a research fellowship.
S.H. is supported by Ministry of Education, Science and
Culture under grant number 09740226.
%%%%%%%%%%%%%%%%%%%%%%% REFERENCES %%%%%%%%%%%%%%%%%%%%%%%%%%%%%%

%%%%%%%%%%%%%%%%%%%% Tables %%%%%%%%%%%%%%%%%%%
\newpage
%%%%%%%%%%%%%% binding 2 %%%%%%%%%%%%%
\begin{table}[H]
\begin{tabular}{ccccc} 
$am_Q$& $\kappa$=0.1570 & 0.1585 & 0.1600 & $\kappa_c$=0.16346 \\ 
                              \hline 
       5.0 & 0.625(7) & 0.602(8) & 0.577(12) & 0.524(16)  \\
           & 0.631(7) & 0.608(8) & 0.583(11) & 0.531(15)  
\vspace{0.1cm} \\ 
       2.6 & 0.618(5) & 0.595(6) & 0.570(8)  & 0.518(11)  \\
           & 0.624(5) & 0.601(6) & 0.576(8)  & 0.524(11)  
\vspace{0.1cm} \\ 
       2.1 & 0.613(5) & 0.590(6) & 0.565(7)  & 0.512(10)  \\
           & 0.618(5) & 0.594(6) & 0.569(7)  & 0.516(10)  
\vspace{0.1cm} \\ 
       1.5 & 0.604(4) & 0.580(5) & 0.555(6)  & 0.501(8)   \\
           & 0.600(4) & 0.576(5) & 0.551(6)  & 0.498(8)  
\vspace{0.1cm} \\ 
       1.2 & 0.596(4) & 0.571(5) & 0.546(6)  & 0.492(8)   \\
           & 0.581(4) & 0.556(5) & 0.531(6)  & 0.477(7)  
\vspace{0.1cm} \\ 
       0.9 & 0.579(4) & 0.554(4) & 0.529(5)  & 0.473(7)  \\ 
           & 0.536(4) & 0.511(4) & 0.486(5)  & 0.430(7)   \\ 
\end{tabular}
\caption{Binding energies of heavy-light pseudoscalar mesons
  for each $\kappa$ and set. Upper lines are from set I and
  lower lines from set II.} 
\label{bind2}
\end{table}
%%%%%%%%%%%% f \sqrt{M} %%%%%%%%%%%%%%%%%%%%
\begin{table}[H]
\begin{tabular}{ccccc}
$am_Q$ & $\kappa$=0.1570 & 0.1585 & 0.1600 & $\kappa_c$=0.16346 
\\  \hline 
5.0 & 0.408(13)& 0.377(13)& 0.341(14)& 0.268(17) \\
           & 0.397(12)& 0.367(12)& 0.333(13)& 0.264(16) 
\vspace{0.1cm} \\ 
       2.6 & 0.334(7) & 0.311(8) & 0.286(8) & 0.233(10) \\
           & 0.323(7) & 0.301(7) & 0.277(8) & 0.227(9)  
\vspace{0.1cm} \\ 
       2.1 & 0.310(6) & 0.289(7) & 0.266(7) & 0.219(8)  \\
           & 0.298(6) & 0.278(6) & 0.257(6) & 0.212(8)  
\vspace{0.1cm} \\ 
       1.5 & 0.269(5) & 0.252(5) & 0.234(5) & 0.195(7)  \\
           & 0.256(5) & 0.240(5) & 0.223(5) & 0.187(7)  
\vspace{0.1cm} \\ 
       1.2 & 0.242(4) & 0.227(4) & 0.212(5) & 0.178(6)  \\
           & 0.224(4) & 0.211(4) & 0.197(5) & 0.168(6)  
\vspace{0.1cm} \\ 
       0.9 & 0.209(4) & 0.197(4) & 0.184(4) & 0.156(5)  \\ 
           & 0.176(4) & 0.166(4) & 0.157(5) & 0.134(6)  \\ 
\end{tabular}
\caption{Numerical results for $(f_P M_P^{1/2})^{latt}$ for
  each $\kappa$ and set. Upper lines are from set I and
  lower lines from set II.}
\label{frootm}
\end{table}
%%%%%%%%%%%%%%%% quadratic fit of fmz_P %%%%%%%%%%%%%%%%
\begin{table}[H]
\begin{tabular}{cccc}
 set        & $(f_P\sqrt{M_P})^{\infty}$ & $a_1$     & $a_2$  \\
 \hline
 set I      & 0.320(22)                  & -0.97(11) & 0.36(10) \\ 
 set II & 0.308(20)                  & -0.87(11) & 0.17(11) \\
\end{tabular}
\caption{The coefficients obtained by fitting each data to a quadratic 
function.}
\label{fit}
\end{table}
%%%%%%%%%%%%%%%% delta fmz_P %%%%%%%%%%%%%%%%
\begin{table}[H]
\begin{tabular}{cccccc}
 $am_Q$              &  $\delta f_P^{(1)}$  &  $\delta f_P^{(2)}$ &
 $\delta f_P^{(3)}$  &  $\delta f_P^{(4)}$  &  $\delta f_P^{(5)}$  \\ 
  &&($\times$ 100)&($\times$ 100)&($\times$ 100)&($\times$ 100)    \\ 
 \hline
 5.0 & 0.290(18)& -2.2(3) &          &         & \\
     & 0.288(16)& -2.1(3) & -0.18(3) & 0.03(1) & -0.066(15) 
\vspace{0.1cm} \\ 
 2.6 & 0.267(11)& -3.3(3) &          &         & \\
     & 0.266(10)& -3.3(3) & -0.46(4) & 0.13(2) & -0.24(3) 
\vspace{0.1cm} \\ 
 2.1 & 0.258(9) & -3.9(3) &          &         & \\
     & 0.259(9) & -3.9(3) & -0.65(5) & 0.21(2) & -0.35(3)
\vspace{0.1cm} \\ 
 1.5 & 0.243(7) & -4.9(3) &          &         & \\            
     & 0.252(7) & -5.1(3) & -1.11(7) & 0.43(4) & -0.69(5)
\vspace{0.1cm} \\ 
 1.2 & 0.235(6) & -5.7(3) &          &         & \\            
     & 0.248(7) & -6.1(4) & -1.59(9) & 0.71(5) & -1.07(7)
\vspace{0.1cm} \\ 
 0.9 & 0.226(6) & -6.9(4) &          &         & \\           
     & 0.247(7) & -7.8(4) & -2.78(10)& 1.35(8) & -1.99(10)\\
\end{tabular}
\caption{Numerical results for $\delta f_P^{(i)}$ at $\kappa
 = \kappa_c$ in lattice unit. Upper lines with set I and lower 
 lines with set II.}
\label{dfmzP}
\end{table}
%%%%%%%%%%%%%%%% fmz_V and \delta fmz_V %%%%%%%%%%%%%%%%
\begin{table}[H]
\begin{tabular}{ccccccc}
 $am_Q$              &  $(f_VM_V^{1/2})^{latt}$ &
 $\delta f_V^{(1)}$  &  $\delta f_V^{(2)}$                 &
 $\delta f_V^{(3)}$  &  $\delta f_V^{(4)}$                 &
 $\delta f_V^{(5)}$   \\ 
 &&&($\times$ 100)&($\times$ 100)&($\times$ 100)&($\times$ 100)
 \\ \hline
 5.0 & 0.280(18)& 0.274(18)& 0.63(11)&          &          & \\
     & 0.275(10)& 0.270(16)& 0.61(10)& -0.16(3) & 0.003(6) & 0.017(5) 
\vspace{0.1cm} \\ 
 2.6 & 0.248(10)& 0.240(9) & 0.87(11)&          &          & \\
     & 0.240(9) & 0.235(9) & 0.86(11)& -0.39(5) & 0.027(9) & 0.06(1) 
\vspace{0.1cm} \\ 
 2.1 & 0.237(9) & 0.227(8) & 1.00(11)&          &          & \\
     & 0.229(8) & 0.223(8) & 0.98(11)& -0.53(6) & 0.05(1)  & 0.09(2)
\vspace{0.1cm} \\ 
 1.5 & 0.220(7) & 0.207(6) & 1.24(12)&          &          & \\
     & 0.234(7) & 0.207(7) & 1.20(13)& -0.83(8) & 0.11(2)  & 0.19(2)
\vspace{0.1cm} \\ 
 1.2 & 0.210(6) & 0.196(6) & 1.42(13)&          &          & \\
     & 0.205(7) & 0.198(6) & 1.36(15)& -1.11(10)& 0.19(3)  & 0.30(4)
\vspace{0.1cm} \\ 
 0.9 & 0.201(6) & 0.184(5) & 1.70(16)&          &          & \\
     & 0.195(7) & 0.188(6) & 1.58(20)& -1.78(16)& 0.37(5)  & 0.55(7)\\
\end{tabular}
\caption{Numerical results for $(f_VM_V^{1/2})^{latt}$ and
 $\delta f_V^{(i)} $ at $\kappa = \kappa_c$ in lattice
 unit. 
 Upper lines with set I and lower lines with set II.}
\label{dfmzV}
\end{table}
%%%%%%%%%%%%%%%% fmz_AV and \delta fmz_AV %%%%%%%%%%%%%%%%
\begin{table}[H]
\begin{tabular}{ccccccc}
 $am_Q$              &  $(fM^{1/2})_{AV}^{latt}$ &
 $\delta f_{AV}^{(1)}$  &  $\delta f_{AV}^{(2)}$            &
 $\delta f_{AV}^{(3)}$  &  $\delta f_{AV}^{(4)}$            &
 $\delta f_{AV}^{(5)}$   \\ 
 &&&($\times$ 100)&($\times$ 100)&($\times$ 100)&($\times$ 100)
 \\ \hline
 5.0 & 0.278(17)& 0.278(18)& -0.06(5) &          &          & \\
     & 0.272(16)& 0.274(16)& -0.07(5) & -0.17(3) & 0.010(4) & -0.004(2)
\vspace{0.1cm} \\ 
 2.6 & 0.244(9) & 0.246(9) & -0.18(6) &          &          & \\
     & 0.237(9) & 0.242(9) & -0.19(6) & -0.41(4) & 0.053(5) & -0.012(6)
\vspace{0.1cm} \\ 
 2.1 & 0.232(8) & 0.235(8) & -0.22(7) &          &          & \\
     & 0.225(8) & 0.232(8) & -0.25(7) & -0.56(5) & 0.089(7) & -0.018(8)
\vspace{0.1cm} \\ 
 1.5 & 0.213(6) & 0.216(7) & -0.30(8) &          &          & \\
     & 0.207(7) & 0.218(7) & -0.37(9) & -0.90(7) & 0.193(14)& -0.032(16)
\vspace{0.1cm} \\ 
 1.2 & 0.202(6) & 0.205(6) & -0.36(9) &          &          & \\
     & 0.195(6) & 0.210(6) & -0.49(11)& -1.23(9) & 0.32(2)  & -0.048(25)
\vspace{0.1cm} \\ 
 0.9 & 0.189(5) & 0.194(5) & -0.46(11)&          &          & \\
     & 0.180(6) & 0.203(6) & -0.76(15)& -2.03(15)& 0.62(4)  & -0.084(51) \\
\end{tabular}
\caption{Numerical results for spin averaged lattice matrix
  elements and their current components at $\kappa =
  \kappa_c$ in lattice unit. 
  Upper lines with set I and lower lines with set II.} 
\label{n-dfmzAV}
\end{table}
%%%%%%%%%%%%%%%% fP/fV %%%%%%%%%%%%%%%%
\begin{table}[H]
\begin{tabular}{ccccccc}
 $am_Q$                   &  $(f_P/f_V)^{latt}$       &
 $\delta (f_P/f_V)^{(1)}$ &  $\delta (f_P/f_V)^{(2)}$ &
 $\delta (f_P/f_V)^{(3)}$ &  $\delta (f_P/f_V)^{(4)}$ &
 $\delta (f_P/f_V)^{(5)}$
 \\ \hline
 5.0 & 0.954(24)& 1.055(20)& -3.42(37)&          &          & \\
     & 0.962(24)& 1.065(18)& -3.47(35)& 1.084(58)& 14(48)   & -3.98(75)
\vspace{0.1cm} \\ 
 2.6 & 0.941(24)& 1.113(16)& -3.83(35)&          &          & \\
     & 0.945(22)& 1.133(16)& -3.89(37)& 1.186(66)& 4.8(2.1) & -3.77(47)
\vspace{0.1cm} \\ 
 2.1 & 0.925(22)& 1.136(16)& -3.89(33)&          &          & \\
     & 0.927(22)& 1.163(17)& -4.00(37)& 1.235(67)& 4.2(1.3) &  -3.76(44)
\vspace{0.1cm} \\ 
 1.5 & 0.887(21)& 1.174(15)& -3.96(31)&          &          & \\
     & 0.877(25)& 1.215(18)& -4.24(39)& 1.335(77)& 3.92(80) & -3.70(41)
\vspace{0.1cm} \\ 
 1.2 & 0.848(22)& 1.198(16)& -4.00(31)&          &          & \\
     & 0.819(27)& 1.253(20)& -4.45(43)& 1.426(87)& 3.80(65) & -3.65(41)
\vspace{0.1cm} \\ 
 0.9 & 0.780(24)& 1.228(17)& -4.09(34)&          &          & \\
     & 0.694(32)& 1.311(24)& -4.92(58)& 1.56(11) & 3.61(53) & -3.61(44)\\
\end{tabular}
\caption{Ratio of pseudoscalar and vector lattice matrix
  elements at $\kappa = \kappa_c$.
Upper lines with set I and lower lines with set II.}
\label{PVtab}
\end{table}
%%%%%%%%%%%%%%%% Simlation results %%%%%%%%%%%%%%%%
\begin{table}[H]
\begin{tabular}{ccrrl}
$am_Q$   &  $aM_P$  
   & $aM_V-aM_P$ & $aM_{P_s}-aM_P$ & $f_{P_s}/f_P$        \\ 
   &&($\times$ 100)&($\times$ 100)&         \\ \hline 
         5.0 & 5.524(16) & 0.81(60) & 5.31(57) & 1.271(35) \\
             & 5.531(15) & 0.86(55) & 5.27(54) & 1.258(32) 
\vspace{0.1cm} \\ 
         2.6 & 3.118(11) & 1.53(53) & 5.26(40) & 1.216(24) \\            
             & 3.124(11) & 1.54(54) & 5.29(39) & 1.214(23) 
\vspace{0.1cm} \\ 
         2.1 & 2.612(10) & 1.80(53) & 5.32(37) & 1.206(21) \\
             & 2.616(10) & 1.84(55) & 5.35(36) & 1.202(21) 
\vspace{0.1cm} \\ 
         1.5 & 2.001(8)  & 2.34(51) & 5.41(33) & 1.185(18) \\            
             & 1.998(8)  & 2.47(56) & 5.41(32) & 1.176(20) 
\vspace{0.1cm} \\ 
         1.2 & 1.692(8)  & 2.81(52) & 5.48(31) & 1.171(17) \\            
             & 1.677(7)  & 3.05(59) & 5.47(30) & 1.159(21) 
\vspace{0.1cm} \\ 
         0.9 & 1.373(7)  & 3.61(55) & 5.57(29) & 1.153(17) \\           
             & 1.330(7)  & 4.06(66) & 5.56(29) & 1.134(24) \\
\end{tabular}
\caption{Simulation results at $\kappa = \kappa_c$ in
   lattice unit. Upper lines with set I and lower lines with set II.} 
\label{num-results}
\end{table}

%%%%%%%%%%%%%%%%%%%%%%%% Figures %%%%%%%%%%%%%%%%%%%%%%%%%%%%

\newlength{\figwidth}
\setlength{\figwidth}{0.65\textwidth}
\addtolength{\figwidth}{-0.5\columnsep}

\newpage
%%%%%%%%% effective plots 1 %%%%%%%%
\begin{figure}
\begin{center}
\leavevmode\psfig{file=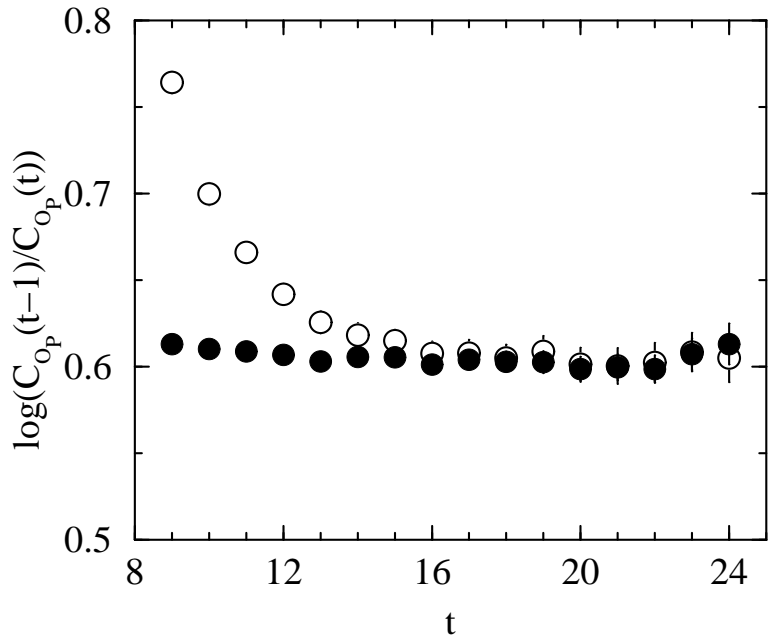,width=\figwidth}
\caption{Effective mass plot with local source(open
  circles) and smeared source (solid circles) at
  $m_Q=2.6,\kappa=0.1585$ with the NRQCD action including
  entire $1/m_{Q}^{2}$ corrections (set II).}
\label{efplt1}
\end{center}
\end{figure}
%%%%%%%%% effective plots 2 %%%%%%%%%
\begin{figure}
\begin{center}
\leavevmode\psfig{file=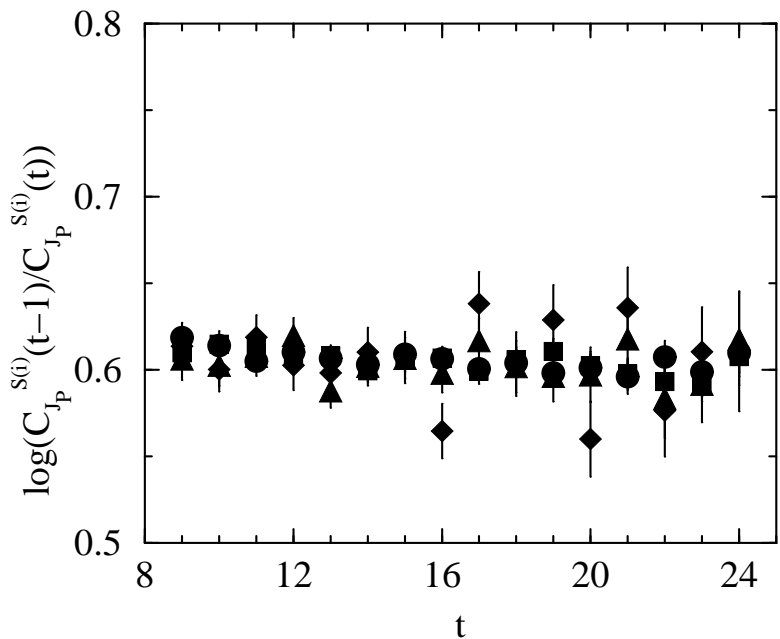,width=\figwidth}
\caption{Effective plot of $C_{J_P}^{S (i)}(t)$ for
  pseudoscalar at $m_Q=2.6,\kappa=0.1585$ with set II. 
  Where i=2,3,4,5 correspond to circles, squares, diamonds
  and triangles, respectively.} 
\label{efplt2}
\end{center}
\end{figure}
%%%%%%%%% chi extrp of mass of PS %%%%%%%%
\begin{figure}[H]
\begin{center}
\leavevmode\psfig{file=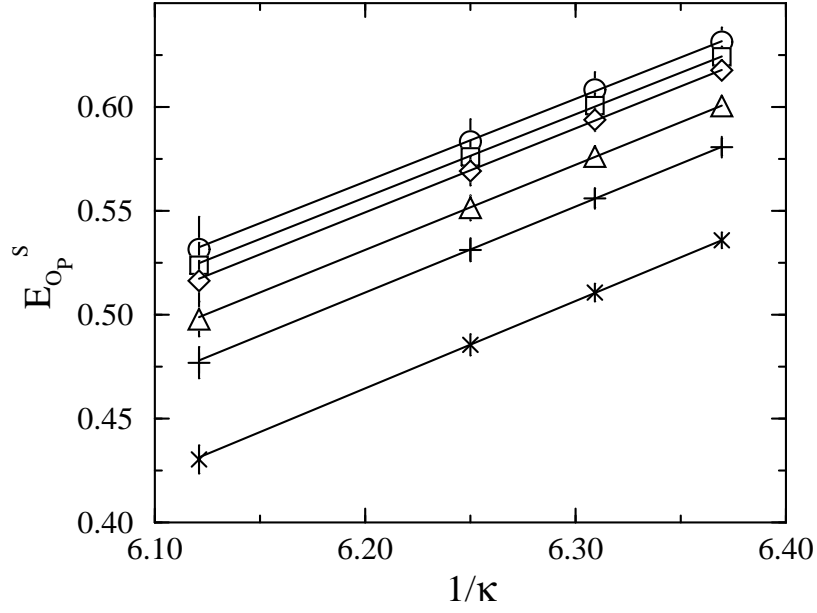,width=\figwidth}
\caption{Chiral extrapolations of of $E_{\Op_P}^S$. 
  From above $am_Q$ = 5.0(circles), 2.6(squares),
  2.1(diamonds), 1.5(triangles), 1.2(pluses) and
  0.9(crosses) in set II.} 
\label{mas-extrp}
\end{center}
\end{figure}
%%%%%%% mass dependence of binding energy %%%%%%%%
\begin{figure}[H]
\begin{center}
\leavevmode\psfig{file=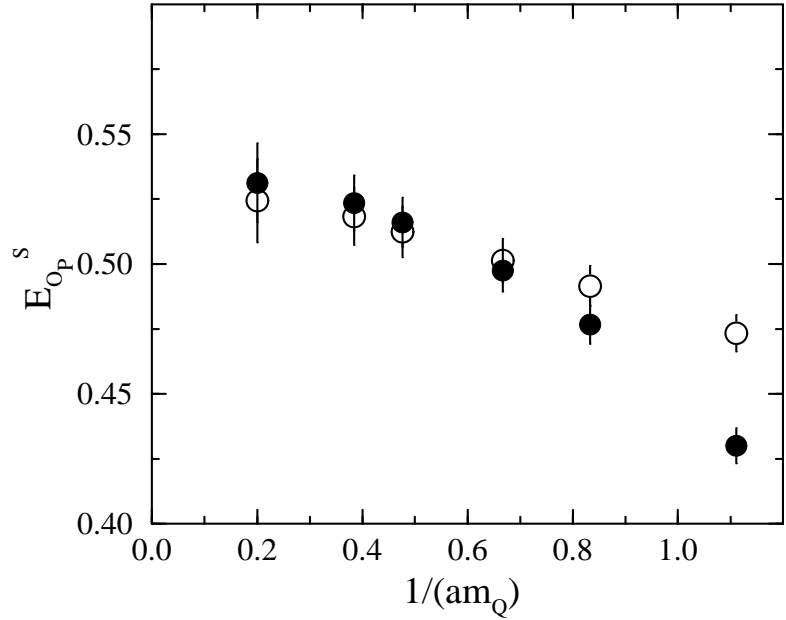,width=\figwidth}
\caption{$1/(am_Q)$ dependence of chirally extrapolated
  binding energies of pseudoscalar(circles) meson 
  from set I(open) and  set II(solid).}
\label{Bind}
\end{center}
\end{figure}
%%%%%%%%% chi-extrp fmz_P %%%%%%%%
\begin{figure}[H]
\begin{center}
\leavevmode\psfig{file=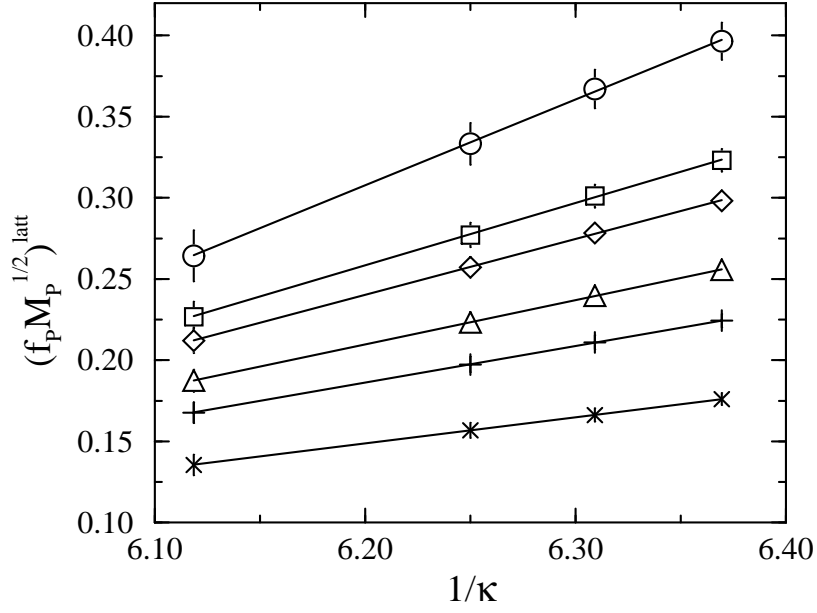,width=\figwidth}
\caption{Chiral extrapolations of $(f_PM_P^{1/2})^{latt}$. 
  From above $am_Q$ = 5.0(circles), 2.6(squares),
  2.1(diamonds), 1.5(triangles), 1.2(pluses) and
  0.9(crosses) in set II. Solid lines are obtained from
  linear fits.} 
\label{chi-frootm}
\end{center}
\end{figure}
%%%%%%%%% fmz_P %%%%%%%%
\begin{figure}[H]
\begin{center}
\leavevmode\psfig{file=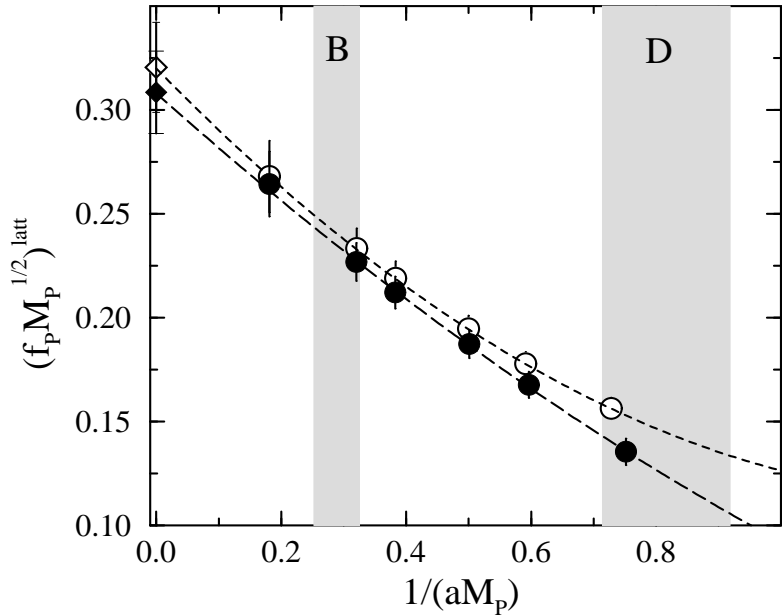,width=\figwidth}
\caption{$1/(aM_P)$ dependence of chirally extrapolated
  $(f_P M_P^{1/2})^{latt}$ with set I(open circles) and
  set II(solid circles). The dashed line is obtained by fitting
  the data from set I to quadratic function and the long dashed line
  from set II.}
\label{fmz}
\end{center}
\end{figure}
%%%%%%%%% fmz_lead %%%%%%%%
\begin{figure}[H]
\begin{center}
\leavevmode\psfig{file=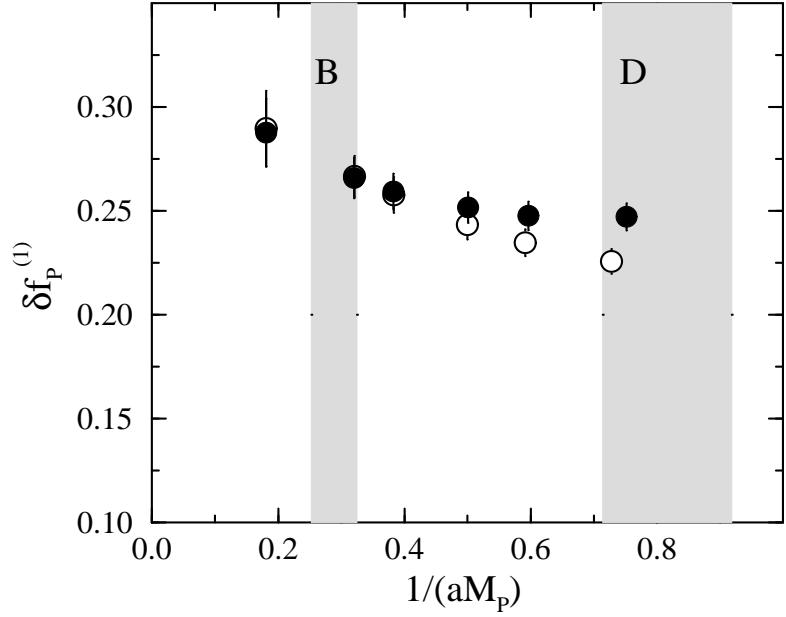,width=\figwidth}
\caption{$1/(aM_P)$ dependence of the leading contribution
  to $(f_P M_P^{1/2})^{latt}$ with set I(open circles) and
  set II(solid circles).}
\label{fmz_lead}
\end{center}
\end{figure}
%%%%%%%%% fmz_other %%%%%%%%
\begin{figure}[H]
\begin{center}
\leavevmode\psfig{file=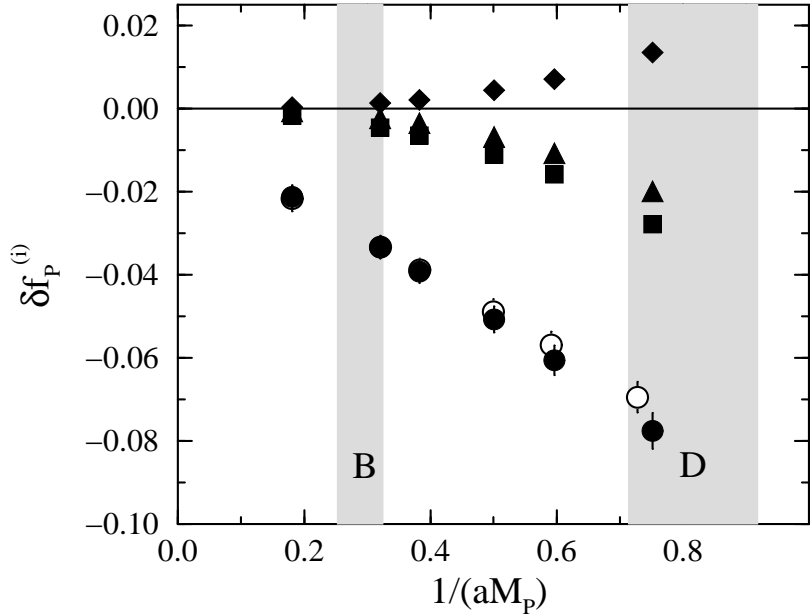,width=\figwidth}
\caption{$1/(aM_P)$ dependence of the non-leading
  contributions to $(f_P M_P^{1/2})^{latt}$ with set I(open
  circles) and set II(solid symbols).
  Solid symbols $\delta f_P^{(i)}$(i=2,3,4,5) are
  corresponding to circles, squares, diamonds and triangles
  respectively.}
\label{fmz_other}
\end{center}
\end{figure}
%%%%%%%%% total correction %%%%%%%%%%%
\begin{figure}[H]
\begin{center}
\leavevmode\psfig{file=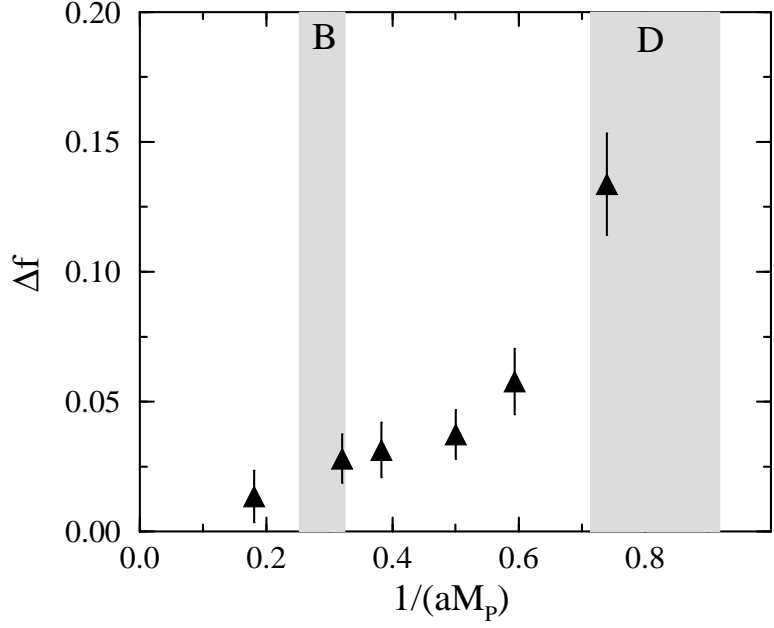,width=\figwidth}
\caption{$1/(aM_P)$ dependence of $O(1/m_Q^2)$ correction.}
\label{total_corr} 
\end{center}
\end{figure}
%%%%%%%%% imaginary total correction %%%%%%%%%%%
\begin{figure}[H]
\begin{center}
\leavevmode\psfig{file=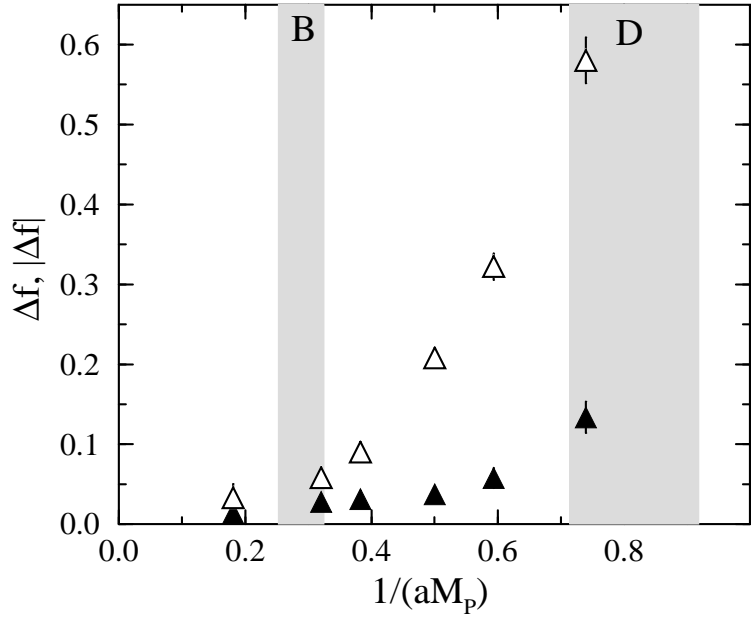,width=\figwidth}
\caption{$1/(aM_P)$ dependence of real $O(1/m_Q^2)$ 
  correction $\tr f$(solid) and imaginary one $|\tr f|$
  (open). For detail, see text.} 
\label{imag_total_corr} 
\end{center}
\end{figure}
%%%%%%%%% fmzAV %%%%%%%%
\begin{figure}
\begin{center}
\leavevmode\psfig{file=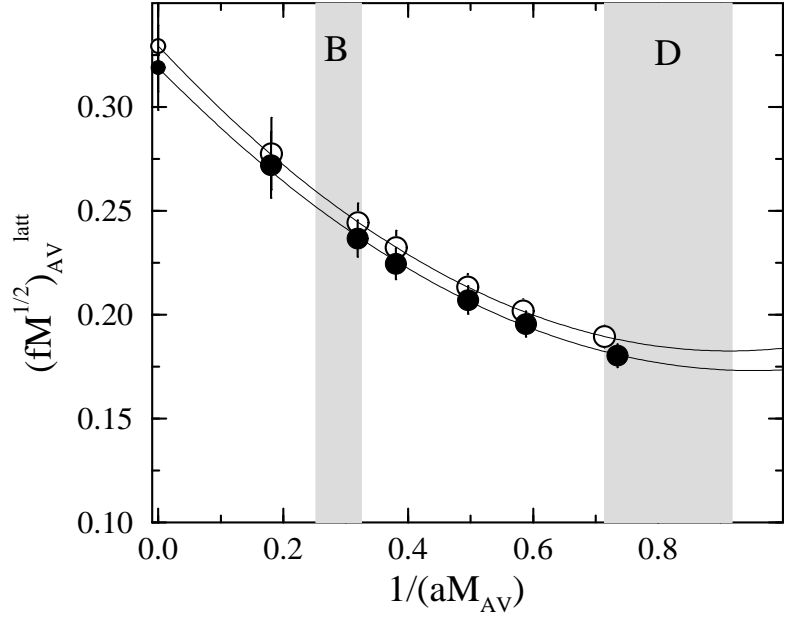,width=\figwidth}
\caption{$1/(aM_{AV})$ dependence of spin averaged
  $(fM^{1/2})_{AV}^{latt}$ with
  set I(open circles) and set II(solid symbols).
Solid line are obtained from a quadratic fit and
Small symbols represent the extrapolated values.}
\label{fmzAV}
\end{center}
\end{figure}
%%%%%%%%% fP/fV %%%%%%%%
\begin{figure}
\begin{center}
\leavevmode\psfig{file=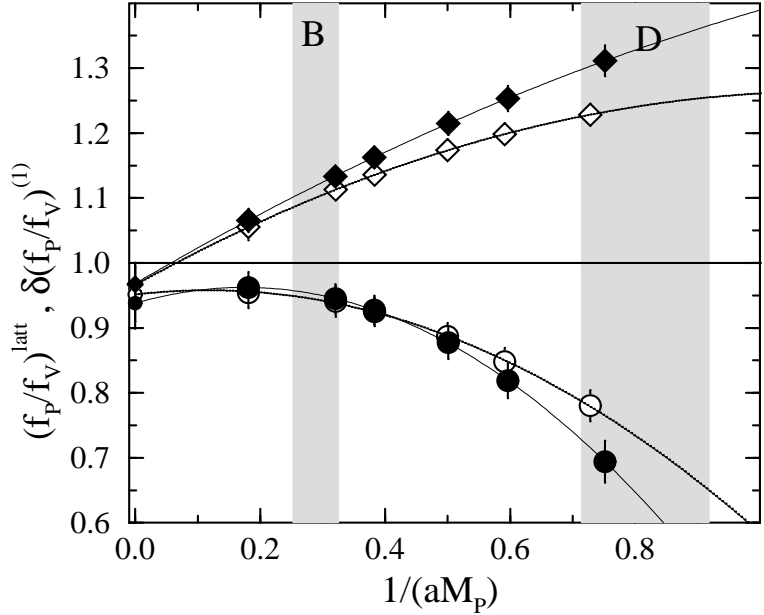,width=\figwidth}
\caption{ $1/(aM_P)$ dependence of the ratio $(f_P/f_V)^{latt}$
and $(f_P/f_V)^{(1)}$ at $\kappa = \kappa_c$
with set I(open symbols) and set II(solid symbols).
Circles refer to $(f_P/f_V)^{latt}$, diamonds refer to 
$(f_P/f_V)^{(1)}$.} 
\label{PVfig}
\end{center}
\end{figure}
%%%%%%%%% 1S hyperfine %%%%%%%%
\begin{figure}[H]
\begin{center}
\leavevmode\psfig{file=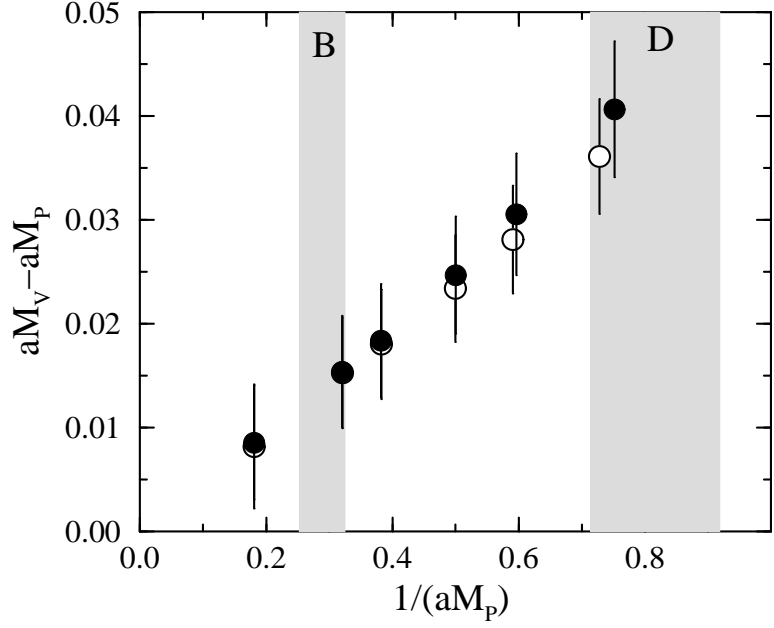,width=\figwidth}
\caption{$1/(aM_P)$ dependence of 1S hyperfine splitting
  with set I(open symbols) and set II(solid symbols).} 
\label{1S-hyper}
\end{center}
\end{figure}
%%%%%%%%% MBs-MB %%%%%%%%
\begin{figure}[H]
\begin{center}
\leavevmode\psfig{file=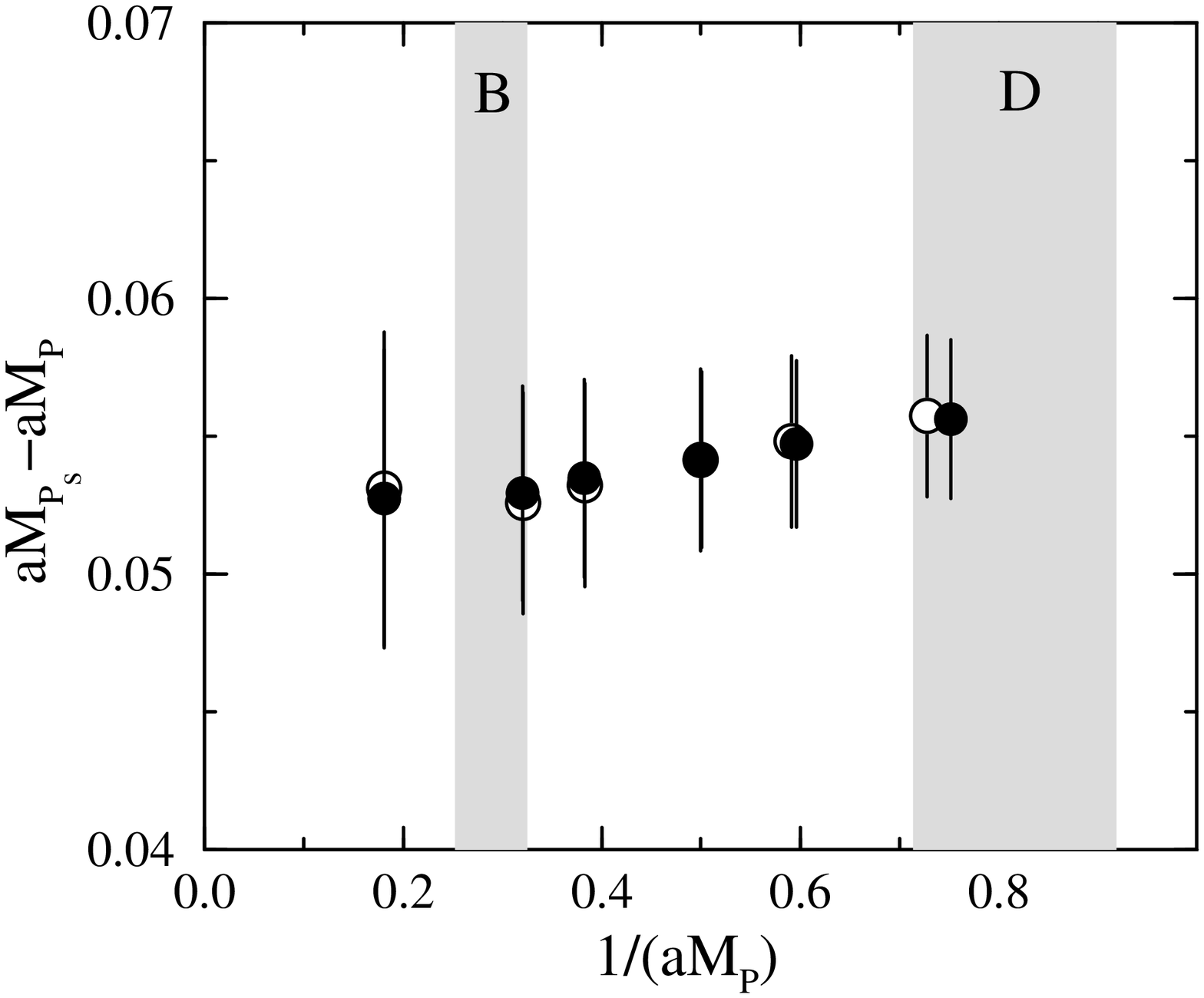,width=\figwidth}
\caption{$1/(aM_P)$ dependence of $M_{P_s}-M_P$ with set
  I(open symbols) and set II(solid symbols).} 
\label{MBs-MB}
\end{center}
\end{figure}
%%%%%%%%% fBs/fB %%%%%%%%
\begin{figure}[H]
\begin{center}
\leavevmode\psfig{file=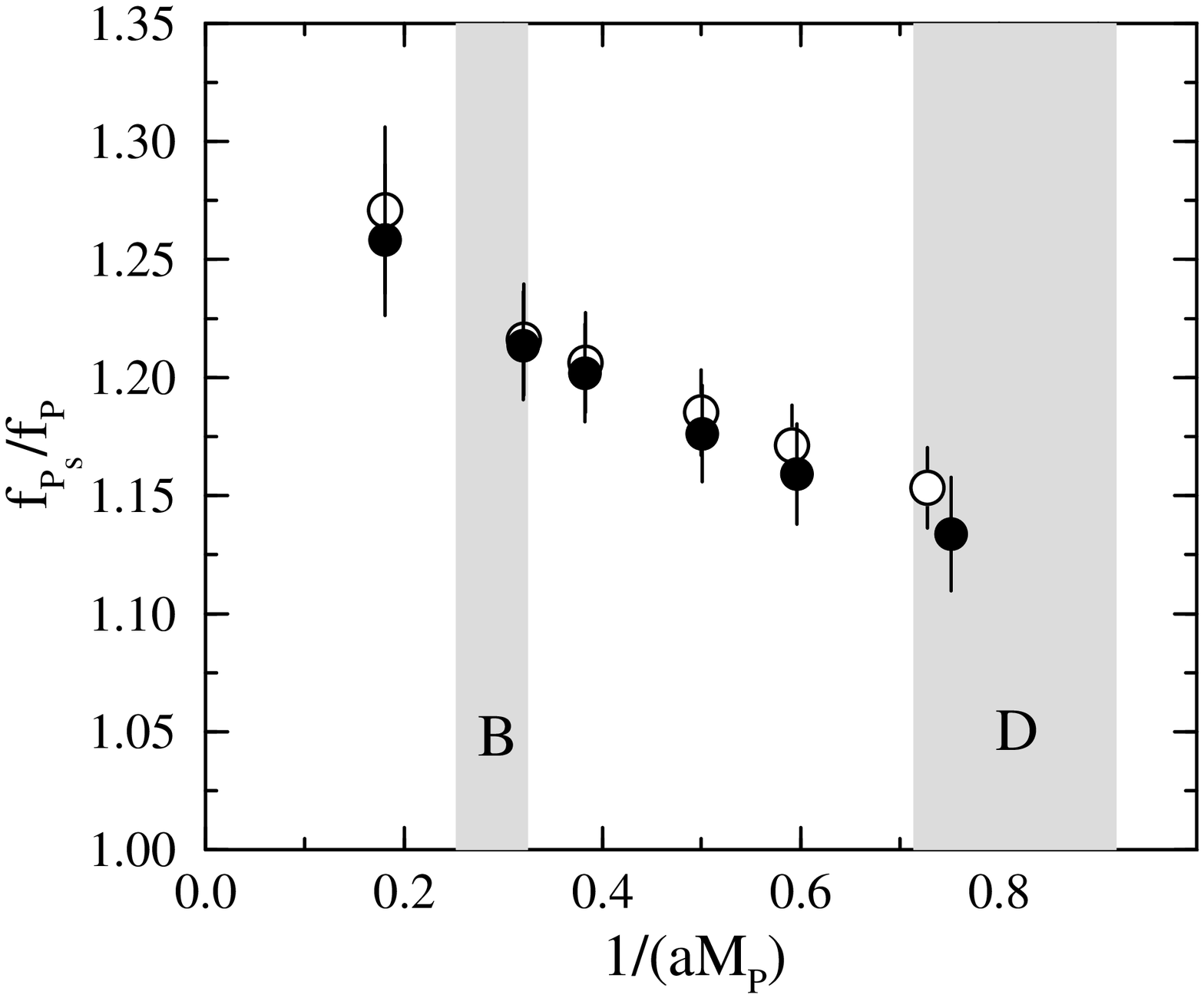,width=\figwidth}
\caption{$1/(aM_P)$ dependence of  $f_{P_s}/f_P$ with set
  I(open symbols) and set II(solid symbols).} 
\label{fBs/fB}
\end{center}
\end{figure}
%%%%%%%%%%%%%%%%%%%%%%%%%%%
%\onecol

%%% Local Variables: 
%%% mode: latex
%%% TeX-master: t
%%% End: 

\end{document}